\documentclass[12pt]{article}
\usepackage{epsfig}
\def\be{\begin{equation}}
\def\ee{\end{equation}}
\def\bea{\begin{eqnarray}}
\def\eea{\end{eqnarray}}
\usepackage{graphicx}

\catcode`\@=11
\def\lsim{\mathrel{\mathpalette\@versim<}}
\def\gsim{\mathrel{\mathpalette\@versim>}}
\def\@versim#1#2{\vcenter{\offinterlineskip
\ialign{$\m@th#1\hfil##\hfil$\crcr#2\crcr\sim\crcr } }}
\catcode`\@=12

\catcode`@=12
\begin{document}
\thispagestyle{empty}
\begin{flushright}
UCRHEP-T544\\
December 2014\
\end{flushright}
\vspace{0.6in}
\begin{center}
{\LARGE \bf SU(2)$_N$ Model of Vector Dark Matter\\ 
with a Leptonic Connection\\}
\vspace{1.2in}
{\bf Sean Fraser, Ernest Ma, and Mohammadreza Zakeri\\}
\vspace{0.2in}
{\sl Department of Physics and Astronomy, University of California,\\
Riverside, California 92521, USA\\}
\end{center}
\vspace{1.2in}
\begin{abstract}\
Models of fermion and scalar dark matter abound.  Here we consider instead 
vector dark matter, from an $SU(2)_N$ extension of the standard model. 
It has a number of interesting properties, including a possible 
implementation of the inverse seesaw mechanism for neutrino mass. 
The annihilation of dark matter for calculating its relic abundance 
in this model is not dominated by its cross section to standard-model 
particles, but rather to other new particles which are in thermal 
equlibrium with those of the standard model.
\end{abstract}

\newpage
\baselineskip 24pt

Whereas the existence of dark matter is universally accepted, its 
nature remains unknown.  It is usually assumed to be a single particle, 
but it may also be more than one~\cite{cmwy07}.  In specific models, it is 
often considered to be a fermion or scalar.  However, vector dark matter is 
certainly also possible~\cite{st03,hm05,h09,dm11,bdmw12,akms12,fr12,bkps13}.
In this paper we consider a variant of an $SU(2)_N$ model proposed 
previously~\cite{dm11,bdmw12}.  The difference is that in our present 
study, all standard-model (SM) fermions are singlets under $SU(2)_N$, whereas 
in the earlier work, that was not the case.

The new particles of our model are the three neutral gauge bosons $X_{1,2,3}$ 
of $SU(2)_N$, three copies of a neutral Dirac fermion $SU(2)_N$ doublet 
$(n_1,n_2)_{L,R}$, a neutral scalar $SU(2)_N$ doublet $(\chi_1,\chi_2)$, and 
a scalar bidoublet $$ \zeta = \pmatrix{\zeta_1^0 & \zeta_2^0 \cr \zeta^-_1 
& \zeta^-_2},$$
which transform (vertically) under $SU(2)_L \times U(1)$ and (horizontally) 
under $SU(2)_N$.  The allowed Yukawa and trilinear scalar couplings are
\begin{eqnarray}
&& (\bar{\nu}_L \zeta_1^0 + \bar{e}_L \zeta_1^-) n_{1R} + 
(\bar{\nu}_L \zeta_2^0 + \bar{e}_L \zeta_2^-) n_{2R}, \\  
&& (\phi^0 \zeta_1^0 - \phi^+ \zeta_1^-) \chi_{1} + 
(\phi^0 \zeta_2^0 - \phi^+ \zeta_2^-) \chi_{2}.
\end{eqnarray}
In analogy to the earlier work~\cite{dm11,bdmw12}, we impose a  
global $U(1)$ symmetry $S'$ on the new particles so that $n$ and $\chi$ 
have $S'=1/2$ and $\zeta$ has $S'=-1/2$.  As $\chi_2$ breaks $SU(2)_N$ 
completely, $T_{3N} + S' = S$ remains exact, under which $n_1,\chi_1 \sim +1$, 
$n_2,\chi_2,\zeta_2 \sim 0$, and $\zeta_1 \sim -1$.  As for the vector gauge 
bosons, $ X (\bar{X}) = (X_1 \mp i X_2)/\sqrt{2} \sim \pm 1$ and 
$Z' = X_3 \sim 0$.  Note that $S'$ 
distinguishes the bidoublet $\zeta$ from its dual $\tilde{\zeta} = \sigma_2 
\zeta^* \sigma_2$, and forbids certain terms in the Lagrangian which would 
be otherwise allowed.  In Refs.~\cite{dm11,bdmw12}, the residual $U(1)$ 
symmetry $S$ is broken explicitly to $(-1)^S$.  Here it remains exact up to 
possible Planck scale corrections, but for the stability of dark matter on 
the scale of the lifetime of the Universe, this is not a serious concern. 
Note that without $\zeta$, the dark sector would only communicate with 
the SM through the Higgs portal.  Here $\zeta$ serves another purpose, i.e. 
neutrino mass.  This connection is of course an assumption of the model, 
but once chosen, it leads naturally to an inverse seesaw mechanism for 
neutrino mass~\cite{ww83,mv86,m87}.  

Consider first the neutrino mass matrix spanning 
$(\bar{\nu}_L, n_{2R}, \bar{n}_{2L})$, i.e.
\begin{equation}
{\cal M}_{\nu n} = \pmatrix{0 & m_D & 0 \cr m_D & 0 & M \cr 0 & M & 0},
\end{equation}
where $m_D$ comes from $\langle \zeta_2^0 \rangle$ and $M$ is an allowed 
invariant mass.  It is clear that this results in one heavy Dirac fermion 
and one massless neutrino corresponding to $\cos \theta \nu_L - \sin \theta 
n_{2L}$, where $\tan \theta = m_D/M$.  As such, lepton number $L$ is 
conserved, with $n_{1,2}$ also having $L=1$  To obtain a nonzero neutrino mass, 
a scalar triplet $\Delta$ under $SU(2)_N$ is required.  Let 
\begin{equation}
\Delta = \pmatrix{\Delta_2/\sqrt{2} & \Delta_3 \cr \Delta_1 & 
-\Delta_2/\sqrt{2}},
\end{equation}
with $S' = -1$, then the terms
\begin{equation}
n_1 n_1 \Delta_1 + (n_1 n_2 + n_2 n_1) \Delta_2/\sqrt{2} - n_2 n_2 \Delta_3,
\end{equation}
where each $n n$ denotes either $n_L n_L$ or $n_R n_R$, 
break $L$ to $(-1)^L$ with a nonzero $\langle \Delta_3 \rangle$, 
without breaking $S$. 
Note that whereas $\langle \Delta_1 \rangle $ as well as $\langle 
\Delta_3 \rangle$ must be nonzero (and large) in the model of 
Ref.~\cite{bdmw12}, thus breaking $S$ to $(-1)^S$, only 
$\langle \Delta_3 \rangle$ is nonzero here and it is very small 
for the implementation of the inverse seesaw~\cite{ww83,mv86,m87}:
\begin{equation}
{\cal M}_{\nu n} = \pmatrix{0 & m_D & 0 \cr m_D & m'_2 & M \cr 0 & M & m_2},
\end{equation}
where $m_2$ comes from $\langle \Delta_3^* \rangle$ and $m'_2$ from $\langle 
\Delta_3 \rangle$.  This means that an inverse seesaw neutrino mass is 
obtained:
\begin{equation}
m_\nu \simeq {m_D^2 m_2 \over M^2}.
\end{equation}

Let $\langle \phi^0 \rangle = v_1$, $\langle \zeta_2^0 \rangle = v_2$, 
$\langle \chi_2 \rangle = u_2$, $\langle \Delta_3 \rangle = u_3$, with 
$g_N$ the $SU(2)_N$ gauge coupling, then the masses of the vector gauge 
bosons of this model are
\begin{eqnarray}
&& m_W^2 = {1 \over 2} g_2^2 (v_1^2 + v_2^2), ~~~~~ 
m_X^2 = {1 \over 2} g_N^2 (u_2^2 + v_2^2 + 2 u_3^2), \\ 
&& m^2_{Z,Z'} = {1 \over 2} \pmatrix{ (g_1^2 + g_2^2)(v_1^2 + v_2^2) & 
-g_N \sqrt{(g_1^2 + g_2^2)} v_2^2 \cr -g_N \sqrt{(g_1^2 + g_2^2)} v_2^2 & 
g_N^2 (u_2^2 + v_2^2 + 4 u_3^2)},
\end{eqnarray}
where $X_3$ has been renamed $Z'$.  In this model, all the SM fermions obtain 
masses from $v_1$, except for the neutrinos which require $v_2$, i.e. $m_D$ 
in Eqs.~(6) and (7).  We assume $v_2$ (which causes $Z-Z'$ mixing) and 
$u_3$ (which breaks $L$ to $(-1)^L$) to be small.  Note that the $Z'$ 
of this model does not couple directly to SM particles, this means that 
it is not easily observable at the Large Hadron Collider (LHC) as other 
$Z'$ bosons which are produced by quarks and decay into lepton pairs.  
This issue will be studied in more detail elsewhere.

Another important 
difference is that whereas the $X_{1,2}$ relic abundance is determind by 
their annihilation cross section to standard-model (SM) particles in 
Ref.~\cite{bdmw12}, it is determined by $X \bar{X} \to \zeta_2 
\zeta_2^\dagger$ here.  Thermalization with SM particles is maintained 
through $\zeta_2 \zeta_2^\dagger \to l^- l^+$, etc.

We now supply the details of this model.  The Higgs potential is given by
\begin{eqnarray}
V &=& \mu_\zeta^2 Tr(\zeta^\dagger \zeta) + \mu_\Phi^2 \Phi^\dagger \Phi + \mu_\chi^2 
\chi^\dagger \chi + \mu_\Delta^2 Tr(\Delta^\dagger \Delta) + (\mu_1 
\tilde{\Phi}^\dagger \zeta \chi + \mu_2 \tilde{\chi}^\dagger \Delta \chi + H.c.)
\nonumber \\ 
&+& {1 \over 2} \lambda_1 [Tr(\zeta^\dagger \zeta)]^2 + {1 \over 2} \lambda_2 
(\Phi^\dagger \Phi)^2 + {1 \over 2} \lambda_3 Tr(\zeta^\dagger \zeta 
\zeta^\dagger \zeta) + {1 \over 2} \lambda_4 (\chi^\dagger \chi)^2 + 
{1 \over 2} \lambda_5 [Tr(\Delta^\dagger \Delta)]^2 \nonumber \\ 
&+& {1 \over 4} \lambda_6 Tr(\Delta^\dagger \Delta - \Delta \Delta^\dagger)^2  
+ f_1 \chi^\dagger \tilde{\zeta}^\dagger \tilde{\zeta} \chi 
+ f_2 \chi^\dagger \zeta^\dagger \zeta \chi 
+ f_3 \Phi^\dagger \zeta \zeta^\dagger \Phi + f_4 \Phi^\dagger 
\tilde{\zeta} \tilde{\zeta}^\dagger \Phi  \nonumber \\ 
&+& f_5 (\Phi^\dagger \Phi)(\chi^\dagger \chi) + 
f_6 (\chi^\dagger \chi) Tr(\Delta^\dagger \Delta) + f_7 \chi^\dagger 
(\Delta \Delta^\dagger - \Delta^\dagger \Delta) \chi + f_8 (\Phi^\dagger \Phi) 
Tr(\Delta^\dagger \Delta) 
\nonumber \\ 
&+& f_9 Tr(\zeta^\dagger \zeta) Tr(\Delta^\dagger \Delta) 
+ f_{10} Tr[\zeta(\Delta^\dagger \Delta - \Delta \Delta^\dagger) \zeta^\dagger],
\end{eqnarray}
where
\begin{equation}
\tilde{\Phi}^\dagger = (\phi^0, -\phi^+), ~~~ \tilde{\chi}^\dagger = 
(\chi_2, -\chi_1), ~~~ \tilde{\zeta} = \pmatrix{\zeta_2^+ & - \zeta_1^+ \cr 
-\bar{\zeta}_2^0 & \bar{\zeta}_1^0}.
\end{equation}
It is the same as that of Ref.~\cite{bdmw12} but with two fewer terms.  
The reason is that $S$ is conserved in our model, whereas $S$ breaks to 
$(-1)^S$ in Ref.~\cite{bdmw12}

The spontaneous symmetry breaking of $SU(2)_N$ is mainly through 
$\langle \chi_2 \rangle = u_2$, where
\begin{equation}
u_2^2 \simeq {- \mu_\chi^2 \over \lambda_4}.
\end{equation}
As previously mentioned, this breaks both $SU(2)_N$ and $S'$, but the 
combination $T_{3N} + S' = S$ remains exact.  The further breaking of $SU(2)_N$ 
by $\langle \Delta_3 \rangle = u_3$ is assumed to be small for the 
implementation of the inverse seesaw mechanism, i.e.
\begin{equation}
u_3 \simeq {-\mu_2 u_2^2 \over \mu_\Delta^2 + (f_6 - f_7) u_2^2}.
\end{equation}
This also does not break $S$, but it breaks $L$ to $(-1)^L$ because of Eq.~(5). 
The spontaneous symmetry breaking of $SU(2)_L \times U(1)_Y$ is mainly through 
$\langle \phi^0 \rangle = v_1$, where
\begin{equation}
v_1^2 \simeq {- \mu_\Phi^2 - f_5 u_2^2 \over \lambda_2}.
\end{equation}
The further breaking of $SU(2)_L \times U(1)_Y \times SU(2)_N$ through 
$\langle \zeta_2^0 \rangle = v_2$ is assumed small, i.e. 
\begin{equation}
v_2 \simeq {-\mu_1 v_1 u_2 \over \mu_\zeta^2 + f_2 u_2^2}.
\end{equation}
This also does not break $S$.

The resulting physical scalar particles of this model have the following 
masses:
\begin{eqnarray}
&& m^2(\sqrt{2} Re \chi_2) \simeq 2 \lambda_4 u_2^2, ~~~ 
m^2(\sqrt{2} Re \phi^0) \simeq 2 \lambda_2 v_1^2, ~~~ 
m^2(\zeta_2^0) \simeq \mu_\zeta^2 + f_2 u_2^2 + f_4 v_1^2, \\ 
&& m^2(\zeta_2^-) \simeq \mu_\zeta^2 + f_2 u_2^2 + f_3 v_1^2, ~~~ 
m^2(\zeta_1^0) \simeq \mu_\zeta^2 + f_1 u_2^2 + f_4 v_1^2, \\ 
&& m^2(\zeta_1^-) \simeq \mu_\zeta^2 + f_1 u_2^2 + f_3 v_1^2, ~~~  
m^2(\Delta_3) \simeq \mu_\Delta^2 + (f_6 - f_7) u_2^2 + f_8 v_1^2, \\  
&& m^2(\Delta_2) \simeq \mu_\Delta^2 + f_6 u_2^2 + f_8 v_1^2, ~~~ 
m^2(\Delta_1) \simeq \mu_\Delta^2 + (f_6 + f_7) u_2^2 + f_8 v_1^2.
\end{eqnarray}
Of all the particles having nonzero $S$, i.e. $\zeta_1^0,\zeta_1^-,\Delta_2,
\Delta_1,n_1$ and $X$, we assume that $X$ is the lightest.  Of all the 
new particles having zero $S$, i.e. $\zeta_2^0,\zeta_2^-,\Delta_3, \sqrt{2} 
Re \chi_2, n_2$ and $Z'$, we assume that $\zeta_2^0,\zeta_2^-$ are lighter 
than $X$ so that $X \bar{X} \to \zeta_2^0 \bar{\zeta}_2^0 + \zeta_2^- \zeta_2^+$ 
is kinematically allowed.  Since $\Delta_1$ has $S=-2$, if $m(\Delta_3) < 
2 m_X$, it is also stable and may become a significant second 
component~\cite{cmwy07} of dark matter.  We will explore this very 
interesting possibility elsewhere.

\begin{figure}[htb]
\vspace*{-3cm}
\hspace*{-3cm}
\includegraphics[scale=1.0]{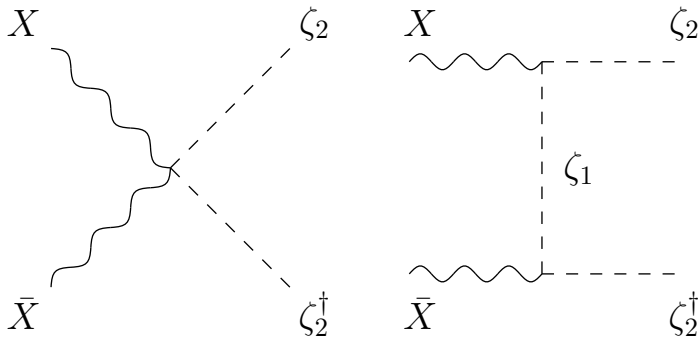}
\vspace*{-20.5cm}
\caption{Annihilation of $X \bar{X}$ to $\zeta_2 \zeta_2^\dagger$.}
\end{figure}
The annihilation of $X \bar{X} \to \zeta_2 \zeta_2^\dagger$ proceeds via the 
diagrams of Fig.~1.  
Assuming $X$ and $\bar{X}$ to be at rest, this 
amplitude is given by
\begin{equation}
{\cal A} = {g_N^2 \over 2} \left[ \vec{\epsilon}_1 \cdot \vec{\epsilon}_2 + 
{4 (\vec{\epsilon}_1 \cdot \vec{k})(\vec{\epsilon}_2 \cdot \vec{k}) \over 
m^2_{\zeta_1} + m_X^2 - m^2_{\zeta_2}} \right],
\end{equation}
where $\vec{k}$ is the three-momentum of $\zeta_2$, and $\epsilon_{1,2}$ 
are the polarizations of $X$ and $\bar{X}$.  
Summing over $\zeta_2^0$ and $\zeta_2^-$, and averaging over the spins of 
$X$ and $\bar{X}$, the corresponding cross section $\times$ their relative 
velocity is given by
\begin{equation}
\sigma \times v_{rel} = {g_N^4 \over 576 \pi m_X^2} \sqrt{1 - {m_{\zeta_2}^2 \over 
m_X^2}} \left( 2 + \left[ 1 + {4(m_X^2 - m_{\zeta_2}^2) \over m^2_{\zeta_1} + 
m_X^2 - m^2_{\zeta_2}} \right]^2 \right).
\end{equation}

\begin{figure}[htb]
\includegraphics[scale=1.5]{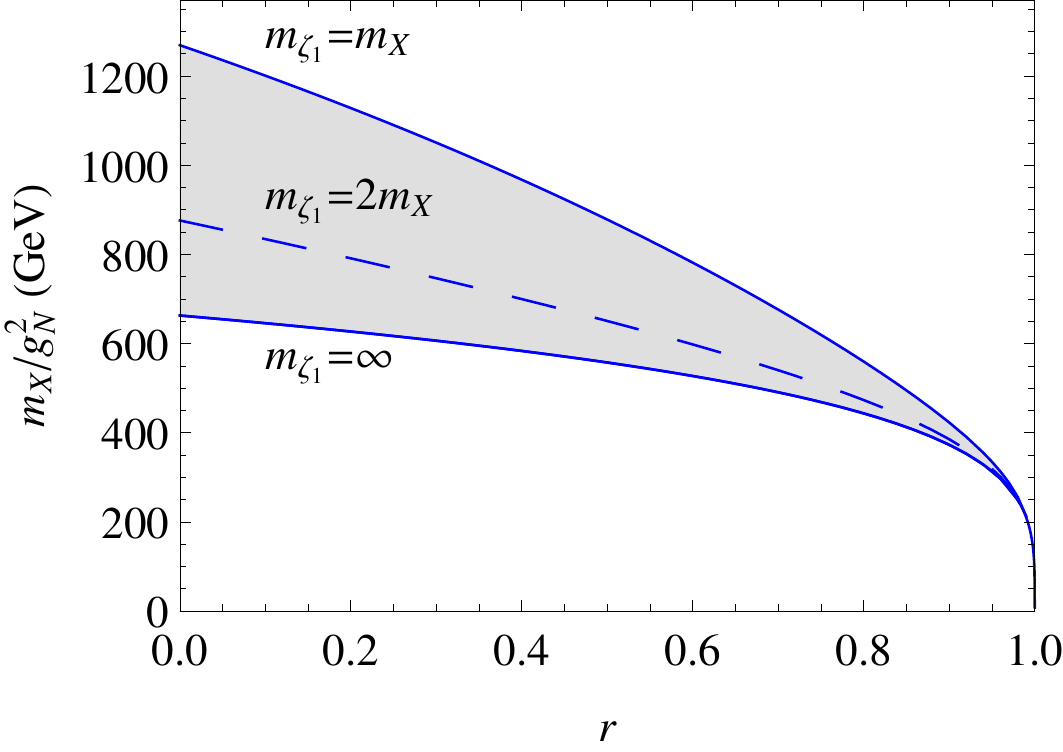}
\caption{Allowed values of $m_X/g_N^2$ plotted against $r=m^2_{\zeta_2}/m_X^2$ 
from relic abundance.}
\end{figure}
Since $m_{\zeta_1} > m_X$ by assumption, the above expression is bounded 
from below by $m_{\zeta_1} \to \infty$ and from above by $m_{\zeta_1} = m_X$. 
We plot in Fig.~2 the allowed region for $m_X/g_N^2$ as a function of 
$r = m^2_{\zeta_2}/m_X^2$ for the optimum value~\cite{sdb12} of 
$\sigma \times v_{rel} = 4.4 \times 10^{-26} ~{\rm cm}^3~{\rm s}^{-1}$ 
implied by the observed relic abundance of dark matter in the Universe, 
taking into account that $X$ is a complex vector field.

To detect $X$ in underground experiments using elastic scattering off 
nuclei, the only possible connection is through $\phi^0 - \chi_2$ mixing. 
The 125 GeV particle $h$ discovered~\cite{atlas12,cms12} at the LHC is 
a linear combination of $\sqrt{2} Re \phi^0$, $\sqrt{2} Re 
\zeta^0_2$, and $\sqrt{2} Re \chi_2$.  The induced $h X \bar{X}$ 
interaction is approximately given by $(g_N^2 v_1/\sqrt{2})(f_5/\lambda_4)$. 
Since $h$ interacts with quarks through $\sqrt{2} Re \phi^0$ according to 
$(m_q/\sqrt{2}v_1) \bar{q} q$, there is a small cross section for $X$ to 
be detected in such experiments.  Following Ref.~\cite{hiny11}, we obtain
\begin{eqnarray}
{f_p \over m_p} &=& -0.075 \left[ {g_N^2(f_5/\lambda_4) \over 4 m_\phi^2} \right] 
- 0.925 (3.51) \left[ {g_N^2(f_5/\lambda_4) \over 54 m_\phi^2} \right], \\ 
{f_n \over m_n} &=& -0.078 \left[ {g_N^2(f_5/\lambda_4) \over 4 m_\phi^2} \right] 
- 0.922 (3.51) \left[ {g_N^2(f_5/\lambda_4) \over 54 m_\phi^2} \right], 
\end{eqnarray}
with the spin-independent elastic cross section for $X$ scattering off a 
nucleus of $Z$ protons and $A-Z$ neutrons normalized to one nucleon given by
\begin{equation}
\sigma_0 = {1 \over \pi} \left( {m_N \over m_X + A m_N} \right)^2 
\left| {Z f_p + (A-Z) f_n \over A} \right|^2.
\end{equation}
Using the recent LUX data~\cite{lux14}, we plot in Fig.~3 the maximum 
allowed value of $g_N^2(f_5/\lambda_4)$ as a function of $m_X$ using 
$m_\phi = 125$ GeV.  We also plot the allowed regions of $g_N^2$ versus 
$m_X$ for $r=0.2$ and $r=0.8$ from Fig.~2.  We see that for moderate 
values of $m_X$, future improvement in direct detection will probe 
the allowed region from relic abundance if $f_5/\lambda_4$ is not too small.
\begin{figure}[htb]
\includegraphics[scale=1.5]{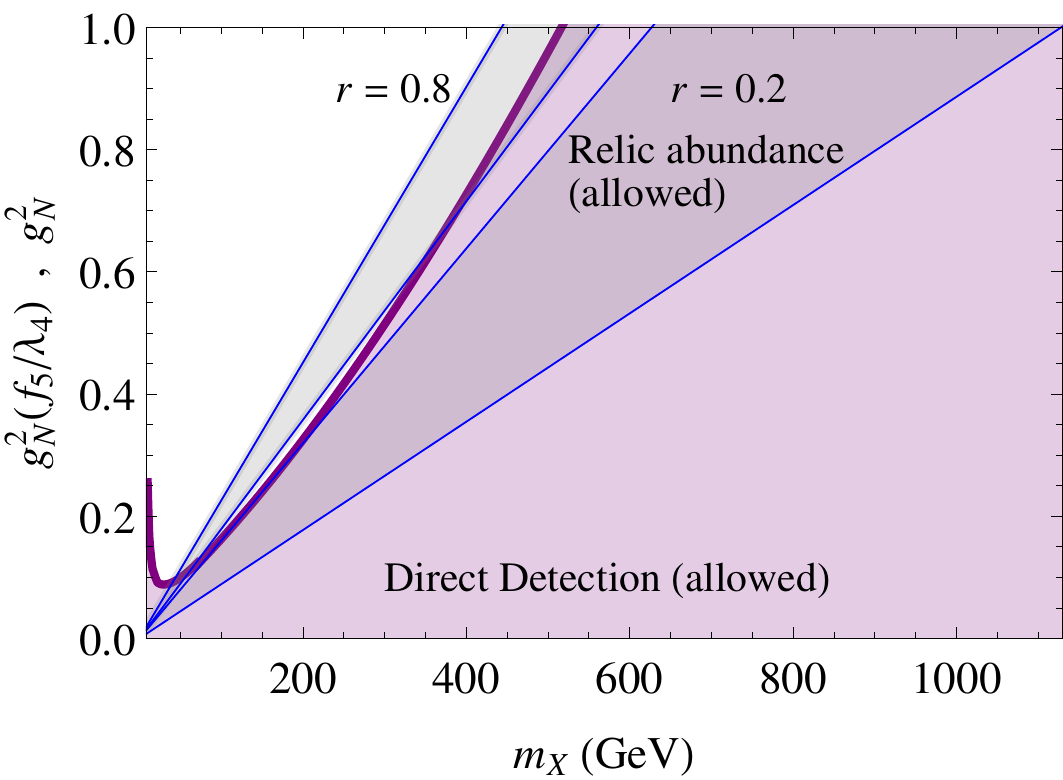}
\caption{Allowed values of $g_N^2$ from relic abundance and 
$g_N^2 (f_5/\lambda_4)$ from LUX, plotted against $m_X$.}
\end{figure}

In conclusion, we have discussed in this paper a simple and realistic model 
of vector dark matter based on $SU(2)_N$.  It has a leptonic connection 
which is a natural framework for the inverse seesaw mechanism to 
generate a small Majorana neutrino mass.  It also accommodates a 
standard-model Higgs boson which mixes only slightly with its 
$SU(2)_N$ counterpart, which does not couple directly to any standard-model 
particle.  Thus the observed 125 GeV particle behaves as the one Higgs boson 
of the standard model for all practical purposes.  The scalar particle 
content of this model may also allow a stable second component of dark matter, 
because of kinematics, and not because the dark symmetry is extended. 
This is the first explicit example of how such a general situation may 
occur, a topic we will explore in another paper.  The $SU(2)_N$ gauge 
symmetry is also suitable to be embedded in an $SU(7)$ model unfifying 
matter and dark matter, as outlined in Ref.~\cite{m13}.

This work is supported in part 
by the U.~S.~Department of Energy under Grant No.~DE-SC0008541.

\bibliographystyle{unsrt}

\end{document}